\documentclass[pra,aps,twocolumn,eqsecnum,showpacs]{revtex4}

\usepackage{dcolumn}
\usepackage{amsmath}
\usepackage{graphics}

\begin{document}

\title{Stimulated Raman process in a scattering medium
in application to quantum memory scheme}

\author{O.S. Mishina${}^{1}$, N.V. Larionov, A.S. Sheremet, I.M. Sokolov, D.V. Kupriyanov}

\affiliation{Department of Theoretical Physics, State Polytechnic
University, 195251, St.-Petersburg, Russia}%

\email{kupr@dk11578.spb.edu}%

\altaffiliation[\\$^{1}$currently at ]{Niels Bohr Institute, University of
Copenhagen, 2100, Copenhagen, Denmark}%

\date{\today }

\begin{abstract}
 We consider the coherent stimulated Raman process developing in an optically
 dense disordered atomic medium, which can also incoherently scatter the light
 over all outward directions. The Raman process is discussed in context of a
 quantum memory scheme and we point out the difference in its physical nature
 from a similar but not identical protocol based on effect of electromagnetically
 induced transparency (EIT). We show that the Raman and EIT memory schemes do not
 compete but complement one another and each of them has certain advantages in
 the area of its applicability. We include in our consideration analysis of
 the transient processes associated with switching off/on the control pulse and
 follow how they modify the probe pulse dynamics on the retrieval stage of the
 memory protocol.
\end{abstract}

\maketitle%


\section{Introduction}

Quantum memories for light pose an extremely important problem for various
protocols of quantum computing and secure communication \cite{BEZ,Ralph,Cerf},
which might be solvable with currently existing technology. At present there
have been several experimental demonstrations of storage and retrieval of
non-classical states of light from macroscopic atomic systems
\cite{KADHP,HAAYANTFK,AYANTFK,JSFCP,NGPSL,CMJLKK,EAMFZ,COBPG,CDLK,HBGHABL}. In
many theoretical discussions the problem was tested in context of achievable
fidelity for storage and retrieval of the original state of light, normally
assumed either coherent, squeezed or single photon
\cite{BEZ,KMP,Cerf,Ralph,SaBPtVL,NWRSWWJ,MKMP}.

In the recently reported series of papers \cite{GALS} thorough analysis was
applied to determine the efficiency of storage and retrieval of a single mode
quantum state of light via its lambda-type conversion into an atomic spin
coherence. Such a conversion is normally associated with either
electromagnetically induced transparency (EIT) two photon resonance, or with
Raman process, or with photon echo effect. Optimization procedure and practical
recommendations are often based on the unique properties of the lambda
transition and presumes the quantum information encoded in collective variables
of the light. For the weak coherent light (with small degeneracy parameter)
this scheme describes, for example, the storage of a single photon quantum bit
and relates the efficiency of the total pulse retrieval to the fidelity of the
single photon quantum state retrieval. Such analysis is relevant for one mode
memory channel and it leaves open important issues concerning storage and
retrieval of a multimode quantum state. The quantum memory for the multimode
configuration in the Raman process has been the subject of recent reports
\cite{NWRSWWJ,MKMP}, which indicated the difficulties and necessity of
additional studies.

In the present paper we shall discuss the stimulated Raman process in a
scattering medium with a $\Lambda$-type excitation, which is initiated by a
control pulse of limited duration. We shall split this mechanism from the
similar but not identical EIT two-photon resonance and discuss the difference
between these processes. The Raman scattering in the Heisenberg formalism,
introduced in Ref.\cite{RaymMost}, and later applied to the quantum memory
problem in Refs.\cite{KMP,NWRSWWJ,MKMP}, is usually discussed for far frequency
offsets of the probe and coupling modes from an atomic resonance and it is
considered in approximation based on adiabatic elimination of the upper states.
This approximation makes possible to build the relevant effective Hamiltonian
and include the losses into the dissipation terms and the Langevin forces. In
the present paper we show how the Raman process can be described for arbitrary
frequency detunings and without such a simplifying approximation. We show how
the Raman process can be discriminated from the EIT two-photon resonance for
any frequency detunings via considering the spectral behavior of the sample
susceptibility dressed by the coupling field.

Our analysis aims to clarify the physical requirements, for dependable
manipulation with the light pulse and atomic coherence at each step of the
complete memory protocol. The important physical processes like atomic motion
and transient effects associated with switching off/on the control pulse will
be described. In the last part of the paper we briefly comment how our approach
can be generalized to a multilevel configuration while paying attention to the
hyperfine structure of alkali atoms, which was earlier considered in a lossless
configuration in Ref.\cite{MKMP}. As an important consequence of our analysis
we shall point out that the EIT resonance and the Raman process are not
competitive protocols but actually complement one another. The protocol based
on the EIT effect works better for pulses longer than the natural atomic
lifetime $\gamma^{-1}$, but the Raman protocol is in principle applicable for
pulses with duration around $\gamma^{-1}$. Thus the Raman scheme can
potentially work for pulses around 20 ns and hence extend the range of pulses,
which could be stored in the atomic spin subsystem. That would lead us to wider
demonstrations of atomic memory for unknown states of light arriving from
radiation sources of different physical nature.

The paper is organized as follows. In section \ref{II} we describe our approach
to the stimulated Raman process, which is based on the Green's function
formalism. We indicate some technical advantages in applying diagram analysis
particularly to the atomic memory problem. In section \ref{III} we make a
diagram derivation of the master equation describing the dynamics of the probe
pulse and atomic coherence on the write-in and retrieval steps of the memory
protocol. Our general discussion is further illustrated by numerical
simulations for the Raman process given in section \ref{IV}. Some details
concerning the Green's function formalism are presented in appendix \ref{A}.

\section{Basic assumptions and calculation approach} \label{II}

Transport of a light pulse in a medium under conditions, where the coherent
forward propagation is damped by losses due to incoherent scattering over all
outward directions, is a quite delicate problem to describe theoretically at
the quantum level. The problem becomes even more subtle for practically
important situation when the energy of the transmitted pulse is comparable with
the scattered light energy. If the losses are significant and the light and
atomic subsystems are initially prepared in arbitrary quantum states it would
be difficult to follow the dynamics of the entire atoms-field subsystem
completely in a three dimensional configuration, see \cite{SornsSorns}. An
essential simplification can be achieved if the light and atomic subsystems are
initially in a coherent state and the interaction process does not modify this
type of state \cite{Coherence}. Such a dynamical process, which reproduces most
of the principle features of the quantum memory protocol, we shall discuss in
this section.

\subsection{Description of the light subsystem}

For a coherent state any quantum characteristics of light can be
defined in terms of the complex amplitude of the electric field given
by the expectation value of its Heisenberg operator considered at
the certain spatial point $\mathbf{r}$, time $t$ and averaged
over the asymptotic state of the system in the infinite past
\begin{equation}
{\cal E}^{(+)}_{\mu}(\mathbf{r},t)=\langle E^{(+)}_{\mu}(\mathbf{r},t)\rangle%
=\langle \mathrm{S}^{\dagger}TE^{(+)}_{0\mu}(\mathbf{r},t)\mathrm{S}\rangle%
\label{2.1}%
\end{equation}
Here $E^{(+)}_{\mu}(\mathbf{r},t)$ and
$E^{(+)}_{0\mu}(\mathbf{r},t)$ are the positive frequency parts of
the $\mu$-th polarization component of the electric field
operators in the Heisenberg and interaction representations
respectively, which are related by the following
unitary transform
\begin{eqnarray}
E^{(+)}_{\mu}(\mathbf{r},t)&=&\mathrm{S}^{\dagger}(t,-\infty)E^{(+)}_{0\mu}(\mathbf{r},t)%
\mathrm{S}(t,-\infty)%
\nonumber\\%
\mathrm{S}(t,t')&=&T\exp(-\frac{i}{\hbar}\int_{t'}^t V_0(t'')dt'')%
\label{2.2}%
\end{eqnarray}
where $V_0(t)$ is the interaction Hamiltonian in interaction
representation, see Eq.(\ref{2.6}) below, $T$ is the time-ordering
operator,  and the S-matrix in expression (\ref{2.1}) is
given by principle limit of the evolutionary operator
$\mathrm{S}=\mathrm{S}(\infty,-\infty)$.

If the light pulse crosses the medium in a single pass, its forward propagation
obeys the mesoscopically averaged Maxwell equation, which can be constituted
and solved under rather general assumptions. The field emerging from the sample
via incoherent scattering can be visualized as secondary waves created by a
disordered system of atomic sources. For the weak field, scattered by the
sample, its intensity, angular and spectral distribution can be found by
analysis of the multiple scattering process. To entirely simulate this process
the Monte-Carlo scheme can be applied.

\begin{figure}[t]
\includegraphics{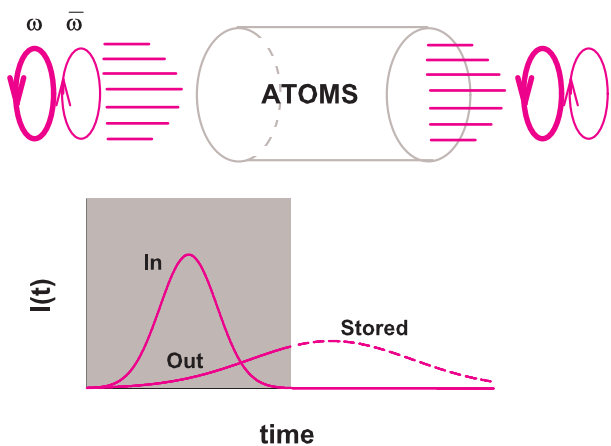}%
\caption{(Color online) Geometry of the quantum memory scheme based on stimulated
Raman process. The left-handed polarized probe pulse with carrier
frequency $\bar{\omega}$ is coherently scattered into the
right-handed polarized strong coupling mode $\omega$. The action
of the coupling field is limited in time how it is indicated by
the shadowed area in the temporal dependence of the process for
input/output probe light intensity $I(t)$. Thus the part of the
probe light will not emerge the sample and will be transformed
into the long-lived spin coherence in atomic subsystem.}
\label{fig1}%
\end{figure}

\begin{figure}[t]
\includegraphics{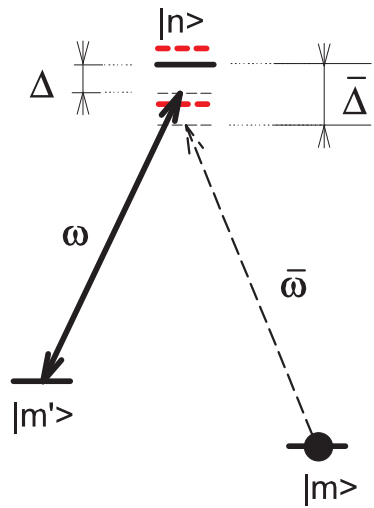}%
\caption{(Color online) Energy diagram illustrating in example of a
$\Lambda$-configured transition how the dressing effects modify
the probe pulse propagation under the stimulated Raman process,
shown in figure \ref{fig1}. The probe pulse with a carrier
frequency $\bar{\omega}$ is scattered into the strong mode of
frequency $\omega$ and creates coherence between the lower Zeeman
states $|m\rangle$ and $|m'\rangle$. In the next step the Zeeman
coherence initiates the delayed retrieval of the probe pulse. If
the coupling mode is "cw" with infinite pulse duration then the
propagation of the probe pulse under the Raman process is damped
by the spontaneous scattering on the components of the
Autler-Townes doublet shown in red-dashed.}
\label{fig2}%
\end{figure}

Let us make these general statements more concrete in context of the
$\Lambda$-configured transition, shown in figures \ref{fig1} and \ref{fig2}.
The depicted scheme is routinely discussed as theoretical background for
description of the stimulated Raman process or EIT two-photon resonance. A
strong coupling field with right-hand circular polarization is quasi-resonantly
applied between a Zeeman hyperfine sublevel of the ground state $|m'\rangle$
and a hyperfine component of the excited level near the upper Zeeman state
$|n\rangle$. A weak probe beam is also quasi-resonant with the $|m\rangle\to
|n\rangle$ Zeeman hyperfine transition. All nonlinear optical effects
associated with the probe light are assumed to be negligible. In our discussion
this field will be taken into account only in the first non-vanishing order.
The atomic medium consists of the atoms initially (in the infinite past)
populating state $|m\rangle$, such that the presence of the weak probe
initiates only weak transfer of atoms from state $|m\rangle$ to $|m'\rangle$.
The sample is optically dense for the probe field and fully transparent for the
coupling field.

Both the coupling and the probe fields are in a coherent state and completely
characterized by the complex amplitude (\ref{2.1}) projected onto the relevant
polarization and/or spectral states that makes these modes distinguishable. The
properties of the coupling field are not modified at all because the sample is
transparent and the number of photons coherently scattered into the coupling
mode is negligible in comparison with the quantum uncertainty of the number of
photons contributing to this mode. Thus the evaluation of the expectation value
(\ref{2.1}) based on a relevant diagram expansion need to be performed only for
the probe field. This calculation will automatically include all the dressing
effects associated with the coupling field and describe in one formalism two
different physical phenomena: stimulated coherent Raman process and EIT
two-photon resonance. The difference between these phenomena will be further
clarified by considering different physical limits and spectral relations
attaining the delay effect and minimizing the incoherent losses.

The global advantage of the coherent state in comparison with
other more sophisticated quantum states, such as squeezed,
entangled etc., is a relatively simple diagram analysis for the
expectation value of the field amplitude (\ref{2.1}). Its
expansion in the series of perturbation theory under application
of Wick's theorem transforms any outcoming product of the field
operators to $N$-ordered form, and generate only the simplest
diagram objects such as the non-perturbed coherent amplitudes of
incoming field and the vacuum-type Green's functions. This makes it
technically possible to further describe the probe field dynamics
in terms of mesoscopically averaged Maxwell equations and multiple
scattering light diffusion.

\subsection{The atomic subsystem and interaction process}

We consider the macroscopic atomic system in the second quantized formalism. In
this subsection we shall show that for an atomic ensemble underwent
coherent-type interaction process shown in figures \ref{fig1} and \ref{fig2}
the quantum dynamics can be properly described by the following four Green's
functions of the Perel-Keldysh diagram technique \cite{Keldysh,LaLfX,BKS}
\begin{equation}
iG_{m'm}^{(\sigma'\sigma)}(\mathbf{r'},t';\mathbf{r},t)=%
\langle T_{\sigma'\sigma}\Psi_{m'}(\mathbf{r'},t')%
\Psi_{m}^{\dagger}(\mathbf{r},t)\rangle%
\label{2.3}%
\end{equation}%
These Green's functions with $\sigma,\sigma'=\mp$ give the expectation values
of the different products of creation and annihilation $\Psi$-operators ordered
by chronological operators $T_{\sigma'\sigma}$. The permutation operators
realize the following action: $T_{--}=T$ arranges the $\Psi$-operators as
ordered in time, its counterpart $T_{++}=\tilde{T}$ arranges them as
anti-ordered in time, operators $T_{+-}$ and $T_{-+}$ are respectively identity
and transposition operators.

The $\Psi$-operators responsible for anihilation or creation of an atom at
spatial point $\mathbf{r}$ and in internal state $|m\rangle$ can be expanded in
the basis set of plane waves as follows
\begin{eqnarray}
\Psi_{m}(\mathbf{r},t)&=&\sum_{\mathbf{p}}\frac{1}{\sqrt{\cal V}}\,%
\mathrm{e}^{\frac{i}{\hbar}\mathbf{p\,r}}\,b_{\mathbf{p}m}(t)%
\nonumber\\%
\Psi_{m}^{\dagger}(\mathbf{r},t)&=&\sum_{\mathbf{p}}\frac{1}{\sqrt{\cal V}}\,%
\mathrm{e}^{-\frac{i}{\hbar}\mathbf{p\,r}}\,b_{\mathbf{p}m}^{\dagger}(t)%
\label{2.4}%
\end{eqnarray}%
where ${\cal V}$ is the quantization volume, and $b_{\mathbf{p}m}(t)$ and
$b_{\mathbf{p}m}^{\dagger}(t)$ are respectively the operators of annihilation
or creation of an atom with momentum $\mathbf{p}$ in internal state $|m\rangle$
in the Heisenberg picture.

If the atomic motion has semiclassical description then the Green's function
$G_{m'm}^{(-+)}(\mathbf{r'},t';\mathbf{r},t)$ before interaction can be
conveniently expressed by the single particle density matrix in the Wigner
representation
\begin{eqnarray}
\lefteqn{iG_{m'm}^{(-+)}(\mathbf{r'},t';\mathbf{r},t)=\pm\int\frac{d^3p}{(2\pi\hbar)^3}\,%
\exp\left[\frac{i}{\hbar}\mathbf{p}(\mathbf{r}'-\mathbf{r})\right]}%
\nonumber\\
&&\times\exp\left[-\frac{i}{\hbar}\frac{p^2}{2\mathrm{m}}(t'-t)-%
\frac{i}{\hbar}\frac{E_{m'}+E_m}{2}(t'-t)\right]%
\nonumber\\
&&\times\rho_{m'm}\left(\mathbf{p},\frac{\mathbf{r}'+\mathbf{r}}{2},\frac{t'+t}{2}\right)%
\label{2.5}%
\end{eqnarray}%
where the external sign $+$ or $-$ corresponds to bosonic or fermionic
statistics of atoms. In this integral relation the exponential factors are
responsible for free dynamics of the atomic state and show fast oscillation in
space and time. The Wigner density matrix $\rho_{m'm}(\mathbf{p},\mathbf{r},t)$
is also oscillating in time with the frequency
$\omega_{m'm}=(E_{m'}-E_{m})/\hbar$ such that the integrand itself has free
precession as function of times $t$ and $t'$ with internal energies $E_m$ and
$E_{m'}$ respectively. The validity of expression (\ref{2.5}) is constrained by
the requirement that the atomic de-Broglie wavelength
$\lambdabar_a\sim\hbar/\sqrt{\mathrm{m}\mathrm{T}}$, where "$\mathrm{m}$" is
atomic mass and $\mathrm{T}$ is ensemble temperature, should be much less than
the scale of spatial inhomogeneity. Then the approximation of the true Green's
function $G_{m'm}^{(-+)}(\mathbf{r'},t';\mathbf{r},t)$ by expression
(\ref{2.5}) is applicable if variations of the increments
$\mathbf{r'}-\mathbf{r}$ and $t'-t$ are small in comparison with the spatial
and time scales where the Wigner function is significantly changed. For free
dynamics and without spin relaxation process the density matrix of the ground
state can only be modified because of atomic motion in free space or in a trap.

It is important to recognize that without interaction with the light subsystem
the Wigner density matrix $\rho_{m'm}(\mathbf{p},\mathbf{r},t)$ gives a
complete description of the macroscopic atomic subsystem. Indeed, as one can
verify each of the remaining Green's functions (\ref{2.3}) can be expressed by
integral relations similar to expression (\ref{2.5}), see \cite{LaLfX,BKS} for
details. The additional terms, which distinguish the four Green's functions,
given by Eq.(\ref{2.3}), are expressed by the commutator (for fermion case by
the anti-commutator) of the free Heisenberg $\Psi$-operators. These terms
contribute to $G^{(+-)}$ with additional multiplication by $\theta(t'-t)$ to
$G^{(--)}$ (retarded type) and by $\theta(t-t')$ to $G^{(++)}$ (advanced type).
In turn each commutator (or anti-commutator) can be expressed by a similar
expansions in the momentum representation and described by similar exponential
functions of $\mathbf{r'}-\mathbf{r}$ and $t'-t$ as in Eq.(\ref{2.5}) if the
atomic motion is free. For a coherent initial state of the atomic system, when
all the atoms occupy only one Zeeman state, any correlation function of higher
order can be expressed as a product of the lower order Green's functions
defined by Eq.(\ref{2.3}).

The coherent type interaction process shown in figure \ref{fig2}
is driven by the following interaction Hamiltonian
\begin{eqnarray}
\!\!\!\!V_0(t)\!&\!=\!&\!-\sum_{n,m}\!\int\! d^3r\,\!(d^\mu)_{nm}%
E^{(+)}_{0\mu}(\mathbf{r},t)\Psi_{0n}^{\dagger}(\mathbf{r},t)\Psi_{0m}(\mathbf{r},t)%
\nonumber\\%
&&\phantom{-\sum_{n,m}\int d^3r\,(d^\mu)_{nm}\,}+ H.c.%
\nonumber\\%
&&E^{(+)}_{0\mu}(\mathbf{r},t)=%
i\frac{\omega_0}{c}A^{(+)}_{0\mu}(\mathbf{r},t)%
\label{2.6}%
\end{eqnarray}%
where all the time dependent operators are written in the interaction
representation as indicated by the zero subscript and $(d^\mu)_{nm}$ denotes
the matrix element of the $\mu$-th component of an atomic dipole. The second
line constitutes the rotating wave approximation and convert to algebraic form
the normal differential relation between the transverse vector potential
operator $A^{(+)}_{0\mu}(\mathbf{r},t)$ and the electric field operator
$E^{(+)}_{0\mu}(\mathbf{r},t)$. This presumes that both the frequencies of the
coupling field $\omega$ and the probe field $\bar{\omega}$ are
indistinguishable from an average frequency $\omega_0$ for the atomic optical
transitions shown in figure \ref{fig2}.

As one can verify via perturbation theory to lowest order the probe field, the
interaction process shown in figure \ref{fig2} will not break the Gaussian
factorization for any higher order correlation functions. This means that the
interaction process leaves the atomic system in the state described by a single
particle density matrix. Any atomic correlation function is described by the
retarded or advanced type Green's functions dressed by interaction with the
strong field. As we shall further show the main optical characteristics of the
sample are expressed by such a dressed retarded Green's function of the excited
state. This Green's function can be found in analytical form for a control
pulse of arbitrary duration and in quite general assumptions, see appendix
\ref{A}. The propagation of the probe pulse through the medium can be
visualized as a polaritonic wave, both for the stimulated Raman and the EIT
processes.

\section{Dynamics of the probe field and atoms for
$\Lambda$-type excitation in the optically thick sample}\label{III}

\subsection{Write-in stage of the memory protocol}

The dynamics of the entire system consisting of the probe field and atoms can
be described via diagram analysis of the atomic and field Green's functions
introduced in the previous section. After a certain regrouping of partial
contributions generated by the expansion of perturbation theory for the
evolutionary operator one obtains the following Dyson equation for the probe
field amplitude
\begin{equation}
\scalebox{1.0}{\includegraphics*{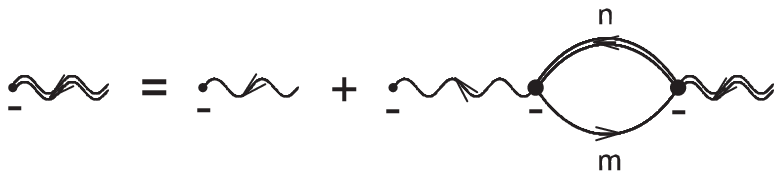}}%
\label{3.1}%
\end{equation}
Here the double wavy single ended lines describe the probe field amplitude
dressed by the coherent interaction shown in figure \ref{fig2}. The single
double ended wavy line is the causal (retarded in the RWA approach) Green's
function for light propagating freely in vacuum. The loop consisted of the
atomic Green's functions and the vertices describe the polarization operator or
susceptibility of the sample in response to the probe field. The equation
(\ref{3.1}) graphically represents the macroscopic Maxwell equation. More
detailed analytical description of the atomic Green's functions is specified in
appendix \ref{A}.

The dynamics of the atomic coherence can be identified if the solution of
equation (\ref{3.1}) is known. This requires evaluation of the following
diagram for the Green's function $G_{m'm}^{(-+)}(\ldots)$ defined by
Eq.(\ref{2.3}) and created by the interaction process
\begin{equation}
\scalebox{1.0}{\includegraphics*{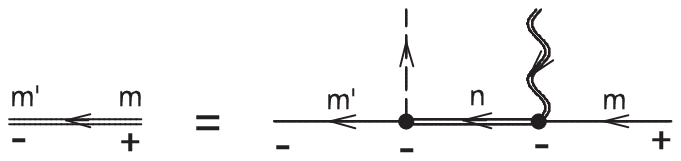}}%
\label{3.2}%
\end{equation}%
Here the internal doubly line represent the retarded type Green's function of
an excited atomic state dressed by the interaction with the vacuum and coupling
modes, see appendix \ref{A}. The outward directed dashed arrow represents the
field amplitude of the coupling mode. The thin lines are related to the
non-perturbed dynamics of free atoms in the ground state and are given by
semiclassical approximation (\ref{2.5}).

The strategic idea of the write-in step of the memory protocol is that equation
(\ref{3.1}) should generate probe pulse delayed so much as it practically does
not emerge from the sample during the action of the control pulse. At the same
time the probe light should not be incoherently scattered by the sample and
after the interaction the information, originally encoded in the probe pulse,
will be concentrated in the spin subsystem. The dynamics of the probe field in
this process is driven by equation (\ref{3.1}), which has the following
analytical form
\begin{equation}
\left[\frac{1}{c}\frac{\partial}{\partial
t}+\frac{\partial}{\partial z}\right]\!%
\epsilon(\mathbf{r}_{\bot},z;t)%
\!=\!2\pi i\frac{\bar{\omega}}{c}\!%
\int\limits_{-\infty}^t\!\!dt'%
\tilde{\chi}(\mathbf{r}_{\bot},z;t,t')\epsilon(\mathbf{r}_{\bot},z;t')%
\label{3.3}%
\end{equation}
where $\epsilon(\mathbf{r}_{\bot},z;t)$ is the slow-varying amplitude of the
probe mode, which is given by a factorization of the probe field component
defined by Eq.(\ref{2.1})
\begin{equation}
{\cal E}^{(+)}_{\mathrm{Left}}(\mathbf{r},t)=%
\epsilon(\mathbf{r}_{\bot},z;t)\mathrm{e}^{-i\bar{\omega}t+i\bar{k}z}%
\label{3.4}%
\end{equation}
The Maxwell equation (\ref{3.3}) indicates that in a dilute sample the coherent
part of the probe pulse propagates only forward and the profile of the outgoing
pulse strongly depends on properties of the sample susceptibility
$\tilde{\chi}(\mathbf{r}_{\bot},z;t,t')$, which controls the efficiency of the
memory protocol. The diffraction divergence was completely ignored in the
derivation of Eq.(\ref{3.3}) and $\mathbf{r}_{\bot}$ was treated only as a
parameter.

Since the coherent process shown in figure \ref{fig2} does not
disturb the original equilibrium distribution of atoms, the ground
state density matrix can be factorized as
\begin{equation}
\rho_{m'm}\left(\mathbf{p},\mathbf{r},t\right)=%
\sigma_{m'm}\left(\mathbf{r}_{\bot},z,t\right)f_0(\mathbf{p})%
\label{3.5}%
\end{equation}
where $f_0(\mathbf{p})$ is the equilibrium Maxwell distribution of the atomic
momenta. In the case of an isolated single $\Lambda$-type excitation the
diagram expression (\ref{3.2}) leads to the following kinetic equation for the
off-diagonal ground state matrix elements of the atomic density matrix
\begin{eqnarray}
\lefteqn{\frac{\partial}{\partial
t}\sigma_{m'm}\left(\mathbf{r}_{\bot},z,t\right)%
+i\omega_{m'm}\,\sigma_{m'm}\left(\mathbf{r}_{\bot},z,t\right)=\Lambda_{m'm}^{n}}%
\nonumber\\%
&&\times\mathrm{e}^{-i(\bar{\omega}-\omega)t+i(\bar{k}-k)z}%
\int\limits_{-\infty}^t\!\!dt'%
\tilde{\chi}(\mathbf{r}_{\bot},z;t,t')\epsilon(\mathbf{r}_{\bot},z;t')\phantom{(3.6)}%
\label{3.6}%
\end{eqnarray}
where the scaling factor
\begin{equation}
\Lambda_{m'm}^{n}=\frac{V_{nm'}^{*}}{\hbar(\mathbf{de}_{-})_{nm}^{*}}%
\label{3.7}%
\end{equation}
is responsible for the difference in the interaction vertices contributing to
the polarization and self-energy operators of the diagrams (\ref{3.1}) and
(\ref{3.2}) respectively. Here $V_{nm'}=(\mathbf{de}_{+})_{nm'}{\cal E}$ is the
transition matrix element for interaction with the strong coupling field, and
$|m'\rangle$, $|m\rangle$ are the Zeeman states coupled by the interaction
process. Vectors $\mathbf{e}_{+}$ and $\mathbf{e}_{-}$ are respectively unit
polarization vectors for right-handed and left-handed polarizations. The strong
mode is assumed to be described by the following positive frequency component
\begin{equation}
{\cal E}^{(+)}_{\mathrm{Right}}(\mathbf{r},t)=%
{\cal E}\,\mathrm{e}^{-i{\omega}t+ikz}\,S(t-z/c)%
\label{3.8}%
\end{equation}
i. e. to be nearly monochromatic during the action of the control
pulse with the envelope profile $S(\tau)$.

Both equations (\ref{3.3}) and (\ref{3.6}) are described by the same sample
susceptibility. The hardest obstacle for further application of equation
(\ref{3.3}) is that the susceptibility is not known analytically in the general
case, when atoms have a multilevel energy structure and the control pulse have
an arbitrary envelope profile. It can be described analytically only after a
round of approximations with respect to the dressing effects associated with
the coupling field and the internal atomic interaction. In appendix \ref{A} we
calculate it for the case of rectangular profile, when the control pulse has
duration $T$
\begin{equation}
S(\tau)=\theta(\tau)-\theta(\tau-T)%
\label{3.9}
\end{equation}
where $\theta(\tau)$ is the step function. For this particular case the Maxwell
equation (\ref{3.3}) can be solved numerically. Then the distribution of the
spin coherence in space and in time can be found via straightforward
integration of equation (\ref{3.6}).

Together the equations (\ref{3.3}) and (\ref{3.6}) describe transport of a spin
polariton mode through the atomic sample, which dynamics can be approximated by
either EIT-type \cite{FleischhLuk} or Raman-type \cite{RaymMost}. As was
mentioned above, the strategic line of the write-in step of the memory protocol
is to stop the control pulse at a time, when the probe mode has not emerged
from the sample and the polariton is localized at least on the scale of the
sample. Then after a round of transient processes the state of the light will
be mapped onto spatially extended state of the atomic spins in the sample.
Without any external disturbances this state will be preserved and will store
the mapped quantum state of the probe light.

In ideal situation this process does not add extra noise to the state. The
crucial requirement to make the memory protocol feasible, as it was discussed
in Refs.\cite{FIM,NWRSWWJ,GALS}, is that the sample should have extremely high
optical thickness, such that $n_0\lambdabar^2L\gg 1$, where $n_0$ is a density
of atoms, $\lambdabar$ is a wavelength divided by $2\pi$ for either probe or
coupling mode, and $L$ is the sample length. If this requirement is fulfilled,
which is principally important for both the Raman and EIT memory schemes, the
delay time for the transport of the probe pulse through the sample can be made
longer than the input pulse duration and the spin polariton can be localized in
the sample.

\subsection{Retrieval stage of the memory protocol}\label{IIIB}

If the coupling light is switched off and after a controllable delay switched
on again, the probe pulse can be stored and retrieved. The delay time is mainly
limited by relaxation processes for the spin coherence. In the case of cold
atoms the spin relaxation can be minimized but atoms can leave the interaction
area because of free motion while the confining trap fields are switched off
during the entire protocol time. However for delay up to few milliseconds the
atoms can be considered as non-drifting and the retrieval stage can be
initiated by recovering the polariton dynamics with the second coherent pulse.

The probe field dynamics is now described by the following Dyson
equation
\begin{widetext}%
\begin{equation}
\scalebox{1.0}{\includegraphics{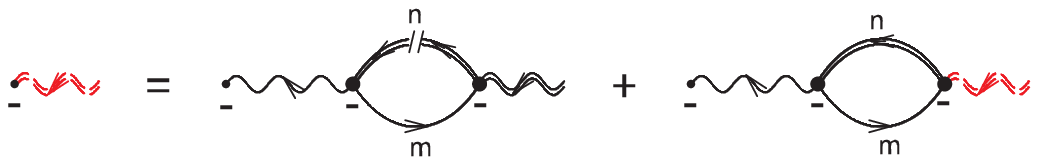}}%
\label{3.10}%
\end{equation}%
\end{widetext}%
The first term on the right hand side displays the source for the optical
coherence initiated by the second control pulse. The black wavy line performs
here electric field of the incoming probe pulse, i. e. the solution of equation
(\ref{3.1}). Since the probe pulse has arrived with the write-in control pulse
the right vertex of this term relates to times associated with the write-in
stage of the protocol. But the left vertex relates to times belonging the
readout control pulse.

The broken double line denotes the retarded Green's function of the excited
atomic state dressed by both the write-in and readout control pulses and has
the following diagram definition
\begin{equation}
\scalebox{1.0}{\includegraphics*{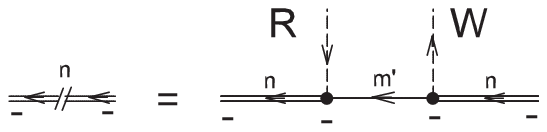}}%
\label{3.11}%
\end{equation}%
The incoming double line is the excited state Green's function dressed by the
write-in control pulse. The outgoing double line is the excited state Green's
function dressed by the readout control pulse. These two lines are converted
into the ground state vacuum line at the vertex points associated with the
coupling fields for write-in (W) or readout (R) control pulses. Just because
the vacuum dynamics, described by the thin solid line, preserves the atomic
state for infinitely long time this diagram being inserted into equation
(\ref{3.10}) can be visualized as the spin coherence, which was originally
created by the process (\ref{3.2}), and which now regenerates the weak probe
field via a stimulated Raman process when the readout control pulse is applied.

The second term in equation (\ref{3.10}) is similar to the second term in
equation (\ref{3.1}) and is responsible for the Maxwell dynamics of the
recovered probe field. The red dashed wavy line reproduces the recovered probe
field for which this equation actually should be solved. In analytical form the
diagram equation (\ref{3.10}) can be written as follows
\begin{eqnarray}
\lefteqn{\left[\frac{1}{c}\frac{\partial}{\partial
t}+\frac{\partial}{\partial z}\right]\!%
\epsilon_{\mathrm{out}}(\mathbf{r}_{\bot},z;t)}%
\nonumber\\%
&&=2\pi i\frac{\bar{\omega}}{c}\!%
\int\limits_{-\infty}^t\!\!dt'%
\tilde{\chi}_{/\!/}(\mathbf{r}_{\bot},z;t,t')\epsilon(\mathbf{r}_{\bot},z;t')%
\nonumber\\%
&&+2\pi i\frac{\bar{\omega}}{c}\!%
\int\limits_{-\infty}^t\!\!dt'%
\tilde{\chi}(\mathbf{r}_{\bot},z;t,t')\epsilon_{\mathrm{out}}(\mathbf{r}_{\bot},z;t')%
\label{3.12}%
\end{eqnarray}%
where $\epsilon(\mathbf{r}_{\bot},z;t')$ is the solution of Eq.(\ref{3.3}) and
$\epsilon_{\mathrm{out}}(\mathbf{r}_{\bot},z;t')$ is field amplitude of the
recovered pulse, for which equation (\ref{3.12}) should be solved. It is
assumed that at the beginning of the readout control pulse the incoming probe
pulse is over. Despite the lower limits of the integrals in the right hand side
are in infinite past, in reality the integrands are non-vanishing only if $t'$
belongs the writing control pulse in the first integral and the readout pulse
in the second. The susceptibilities $\tilde{\chi}_{/\!/}(\ldots)$ and
$\tilde{\chi}(\ldots)$ are given in appendix \ref{A} in example of an isolated
$\Lambda$-type transition.

The dynamics of atomic coherence is now described by the following diagram
equation
\begin{widetext}
\begin{equation}
\scalebox{1.0}{\includegraphics{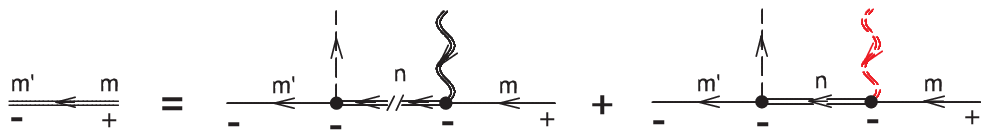}}%
\label{3.13}%
\end{equation}%
\end{widetext}
which in turn leads to the following kinetic equation for the density matrix
(\ref{3.5})
\begin{eqnarray}
\lefteqn{\frac{\partial}{\partial
t}\sigma_{m'm}\left(\mathbf{r}_{\bot},z,t\right)%
+i\omega_{m'm}\,\sigma_{m'm}\left(\mathbf{r}_{\bot},z,t\right)=\Lambda_{m'm}^{n}}%
\nonumber\\%
&&\times\mathrm{e}^{-i(\bar{\omega}-\omega)t+i(\bar{k}-k)z}%
\left[\int\limits_{-\infty}^t\!\!dt'%
\tilde{\chi}_{/\!/}(\mathbf{r}_{\bot},z;t,t')\epsilon(\mathbf{r}_{\bot},z;t')\right.%
\nonumber\\%
&&\left.+\int\limits_{-\infty}^t\!\!dt'%
\tilde{\chi}(\mathbf{r}_{\bot},z;t,t')\epsilon_{\mathrm{out}}(\mathbf{r}_{\bot},z;t')\right]%
\label{3.14}%
\end{eqnarray}
Similar comments to those given for (\ref{3.12}) above apply to this equation,
which is responsible for atomic dynamics at the retrieval stage. In application
it is only necessary to solve equations (\ref{3.3}) and (\ref{3.12}), and then
$\sigma_{m'm}(\mathbf{r}_{\bot},z,t)$ can be found by straightforward
integration of (\ref{3.14}). In practice it is even not so important to know
the solution for remaining atomic coherence because all the retrieved
information goes out with the recovered probe pulse and the equation
(\ref{3.14}) should lead to a trivial result in the infinite future
$\sigma_{m'm}(\mathbf{r}_{\bot},z,t\to\infty)\,\to\,0$. However we introduce
this equation for completeness of the description of the memory protocol.

\section{Results and discussion}\label{IV}

\subsection{The EIT effect and Raman process}

Let us briefly discuss the difference between these two physical mechanisms
when the propagation of the probe field is controlled by a stationary coupling
field. Referring to the definitions of figure \ref{fig2} one can define the
following frequency detunings for the coupling mode
$\Delta=\omega-\omega_{nm'}$ and for the probe mode
$\bar{\Delta}=\bar{\omega}-\omega_{nm}$ from the atomic transition frequencies
$\omega_{nm'}$ and $\omega_{nm}$ respectively. Then one can select two
important resonance relations between the carrier frequency of the probe mode
$\bar{\omega}$ and the energy spectrum of the "dressed" atom. The resonance
point $\Delta=\bar{\Delta}$ defines the two photon resonance, which is
associated with EIT. Indeed the propagation function of the atomic excited
state (\ref{a.3}) as well as the sample susceptibility (\ref{a.13}) vanish at
this point for a pure monochromatic $\Lambda$-type interaction of the modes
$\omega$ and $\bar{\omega}$. The sample becomes transparent and any spectrally
narrow probe pulse distributed in the vicinity of this point will cross the
sample with extremely low group velocity. The light delay, associated with the
EIT effect, does not principally depend on frequency offset $\Delta$ and, in
particular, the group velocity near the EIT point depends only on relation
between the Rabi frequency of the coupling mode, atomic density and $\gamma$
and not on $\Delta$, see Ref.\cite{FIM}. For a sample with high optical depth,
such that $n_0\lambdabar^2L\gg 1$, and for any $\Delta$ the probe pulse
characteristics can always be adjusted to realize its slow transport or storage
in the sample.

Another resonance situation takes place if the carrier frequency of the probe
pulse is scanned near the frequency shifted quasi-energy component of the
Autler-Townes doublet as shown in figure \ref{fig2}. This resonance point
manifests itself as a narrow high amplitude spectral feature in the behavior of
the dispersion and absorption components of the sample susceptibility, see
Ref.\cite{FIM}. In the Maxwell equation (\ref{3.3}) the spectral overlap of the
probe pulse with the dispersion part of the Autler-Townes resonance is
responsible for lossless coherent forward propagation, which is normally
associated with the stimulated Raman process. The overlap with the absorption
part generates the incoherent losses caused by quantum radiation coupling of
such an exciton-type atomic state with the continuum of vacuum modes.

As can be verified in the limit when the Rabi frequency
$\Omega_c=2|V_{nm'}|/\hbar$ is much less than detuning $\Delta$, the equations
(\ref{3.3}) and (\ref{3.6}) can be transformed to the system of equations
introduced in Ref.\cite{RaymMost} and discussed later in the context of quantum
memory in Refs.\cite{KMP,NWRSWWJ,MKMP}. This transformation is equivalent to
the standard procedure of adiabatic elimination of the optical atomic
coherence, which was a basic assumption of the above papers. But we can point
out that such an approximation is only valid if in the integrand in
Eq.(\ref{3.3}) the carrier frequency of the probe mode is located near the
Autler-Townes resonance peak, which always requires for large $\Delta$ some
overlap of the pulse spectrum with this resonance. The importance of this
overlap for the problem of optimization the probe pulse storage was earlier
commented in Ref.\cite{GALS} and in particular we address the reader to the
second paper of that series. It is also important that in the coherent
stimulated Raman process the photons of the probe mode transform to the
coupling mode in result of interaction with all atoms of the ensemble. After a
certain delay the photons of the probe mode are recovered via reversing
interaction of atomic coherence with the coupling field. The Raman mechanism
would exist even if the strong coupling field were exactly on resonance with
the atomic transition, i. e. for $\Delta=0$ when the problem becomes symmetric
with respect to both components of the Autler-Townes doublet. We demonstrate
this possibility by results of our numerical simulations given below.

Applying the Raman process at any $\Delta$ the spontaneous losses can be always
minimized by appropriate adjustment of the external parameters and by
optimization of the pulse shapes, see Refs.\cite{GALS,NWRSWWJ}. This normally
assumes either optimization of the probe pulse parameters with respect to a
given shape and the frequency detuning of the control pulse or alternatively
the optimization of the control pulse with respect to a given probe pulse. In
an idealized lossless situation the stimulated Raman process could be
considered as evolving purely dynamically. But because of the losses the
efficiency of the coherent forward scattering is not responsibly high. In
contrast to the EIT transparency point, for the Raman situation the forward
coherent scattering is more sensitive to the frequency offset $\Delta$ because
the spectral behavior of the sample susceptibility near the Autler-Townes
resonances is essentially modified with varying $\Delta$. Still at a given
detuning and high optical depth the parameters of the probe pulse can be
adjusted for its optimal storage.

Both the Raman and EIT mechanisms for delay of the probe pulse are generally
applicable for any detuning of the coupling mode from the atomic resonance. As
commented above, in a stationary situation, both are described by the spectral
behavior of the same sample susceptibility but considered in different spectral
domains. For general relation among external parameters these spectral domains
can be only conventionally discriminated. However in many practical
realizations the EIT zone and the Raman zone can be clear separated. It is
important to recognize that for both the processes the crucial feature is the
spectral behavior of the dispersion part of the sample susceptibility. It can
be steep either near the transparency point (EIT) or near the absorption point
of the Autler-Townes resonance (Raman).

The delay effect attains by the spectral overlap of the pulse spectrum with the
sample susceptibility near any of these points. For the EIT scheme it is
preferable to keep the pulse spectrum closer to transparency window and the
protocol would work better for longer pulses (in scale of natural lifetime) and
for smaller Rabi frequencies of the coupling field. For Raman scheme the pulse
spectrum should be concentrated close but outward from the absorption peak.
Thus the Raman protocol can be designed for large Rabi frequencies and for
shorter pulses, which duration can be comparable with atomic natural lifetime.
Rigorously speaking the EIT scheme can be also applied for storing of the short
pulses if the Rabi frequency is large such that $\Omega_c\gg\gamma$. But in
this case the slope of the dispersion curve at the EIT point is always  less
steep than in vicinity of the Autler-Townes resonance. Thus the delay effect
would manifest itself more effectively for the pulse which spectrum is located
closer to the Autler-Townes resonance point.

\subsection{Dynamics of the probe pulse under the Raman process}

The general formalism developed in the previous section can be illustrated by
the propagation of a probe pulse with a Gaussian intensity profile
\begin{eqnarray}
\lefteqn{|\epsilon_{\mathrm{in}}(t)|^2\propto
I_{\mathrm{in}}(t)\equiv|\alpha(t)|^2}%
\nonumber\\%
&&=\frac{(\Delta\Omega)}{(2\pi)^{1/2}}%
\exp\left[-\frac{1}{2}\Delta\Omega^2\left(t-\frac{T}{2}\right)^2\right]%
\label{4.1}%
\end{eqnarray}
where for the sake of convenience we rescaled the field amplitude to unit
"pulse energy". The input field amplitude $\alpha(t)$ has the following Fourier
component
\begin{equation}
\alpha_{\Omega}=\frac{(2\pi)^{1/4}}{(\Delta\Omega/2)^{1/2}}%
\exp\left[i\frac{\Omega T}{2}-\frac{\Omega^2}{\Delta\Omega^2}\right]%
\label{4.2}%
\end{equation}
The time shift $T/2$ can either be associated with the central point of the
control pulse or, in a general situation, just defines the time of the probe
pulse arrival. The time or spectral extension of the pulse profile is described
by its spectral variance $(\Delta\Omega/2)^{2}$.

As an illustrative example of the $\Lambda$-configured transition we consider
the Zeeman states in the hyperfine manifold of cesium atoms. As an initial
coherent state, which can be prepared by optical pumping repopulation
processes, we consider the central Zeeman state of the lower (ground) hyperfine
sublevel $|m\rangle=|F_0=3,M_0=0\rangle$, which is a typical situation for
experiments with cesium atomic clocks. The upper state $|n\rangle$, see figure
\ref{fig2}, we will associate with the relevant Zeeman state of the lower
hyperfine sublevel in the excited state of the $D_1$-line, such that
$|n\rangle=|F=3,M=-1\rangle$. The state $|m'\rangle$, coherently coupled to
$|m\rangle$, is the relevant Zeeman state of the upper (ground) hyperfine
sublevel, such that for cesium $|m'\rangle=|F_0=4,M_0=-2\rangle$. This
convention in selection of atomic transitions lets us define the scaling factor
(\ref{3.7}), but all the qualitative results, discussed below, will be
generally valid for any $\Lambda$-scheme. We can also point out that the
considered configuration of atomic transitions is well approximated by
$\Lambda$-scheme up to detunings $\Delta$ more than hundred $MHz$, since the
nearest hyperfine sublevel in the upper state is separated by $1168\, MHz$.

\begin{figure}[t]
\includegraphics{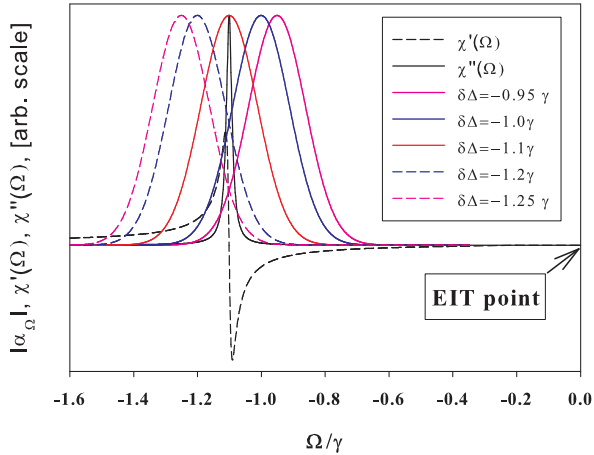}%
\caption{(Color online) Spectral overlap of the the probe pulse amplitude
$\alpha_\Omega$ with the sample susceptibility
$\tilde{\chi}(\Omega)=\tilde{\chi}'(\Omega)+i\tilde{\chi}''(\Omega)$
near the frequency shifted quasi-energy component of the
Autler-Townes doublet, see figure \ref{fig2}. The overlap is
controlled by mismatching the carrier frequencies of the probe and
coupling modes $\delta\Delta=\bar{\Delta}-\Delta$. The graphs are
plotted for the following parameters: $\Delta=-50\gamma$,
$T=50\gamma^{-1}$, $\Delta\Omega=2\pi/T$, $\Omega_{c}=15\gamma$.
The spectra are shifted such that the origin of the plot indicates
the reference point of the EIT two photon resonance.}
\label{fig3}%
\end{figure}

In figure \ref{fig3} we show how the spectral profile $\alpha_{\Omega}$
overlaps with the real and imaginary parts of the susceptibility spectral
component $\tilde{\chi}(\Omega)=\tilde{\chi}'(\Omega)+i\tilde{\chi}''(\Omega)$
when the atoms are dressed by a stationary monochromatic coupling field. The
overlap of the pulse spectrum with the sample susceptibility can be controlled
by mismatching the carrier frequencies/detunings of the coupling and probe
modes $\delta\Delta=\bar{\Delta}-\Delta$, see figure \ref{fig2}. The origin of
the plot in figure \ref{fig3} indicates the location of the EIT point and for
the given spectra the probe pulses do not overlap at this point.

\begin{figure}[t]
\includegraphics{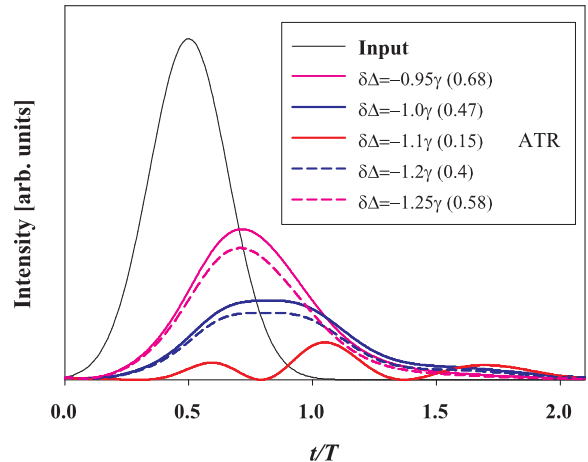}%
\caption{(Color online) Time dependence of the output pulses with the same
initial spectral characteristics as in figure \ref{fig3}, which
have passed the optically thick sample with $n_0\lambdabar^2L=25$.
The numbers in the brackets
indicate the transmittance efficiency. The red curve corresponds
to coincidence of the pulse carrier frequency with the
Autler-Townes resonance and its oscillations indicate
the spectral splitting of the pulse by this resonance.}
\label{fig4}%
\end{figure}

Variation of the spectral location of the probe pulse near the Autler-Townes
resonance leads to different overlaps with the real and imaginary parts of the
sample susceptibility. As a consequence, the influence of absorption
(incoherent scattering) and delay effects on the pulse propagation will be also
different with variation of $\delta\Delta$. The negative role of absorption
essentially falls off with deviation of $\delta\Delta$ from the exact resonance
point (red curve in figure \ref{fig3}), but the positive tendency associated
with pulse delay can survive. This is illustrated in figure \ref{fig4} by
tracking the time dependence of the outgoing pulses, which had the same initial
spectral characteristics as in figure \ref{fig3} and passed through the
optically thick sample with $n_0\lambdabar^2 L=25$. As follows from the plotted
dependencies, when absorption is more or less negligible and the transmittance
efficiency is around 70\%, the pulse delay is still a perceptible effect. Let
us point out that this predicted transmittance efficiency for delayed pulse is
a certain indicator that the Raman protocol in the discussed conditions,
coordinated with the currently existing experimental capabilities, can be
applied, for example, for quantum storage of a single photon state with unknown
polarization \cite{CDLK}. In this situation it overcomes the best classical
estimate "2/3" , see Ref.\cite{MasPop} for detail. It is also noteworthy that
in the perfect resonance situation the outgoing pulse splits into a double-bell
spectral profile, which in turn leads to oscillating time dependence of
intensity, see figure \ref{fig4}.

\subsection{The pulse storage and retrieval}

\begin{figure}[t]
\includegraphics{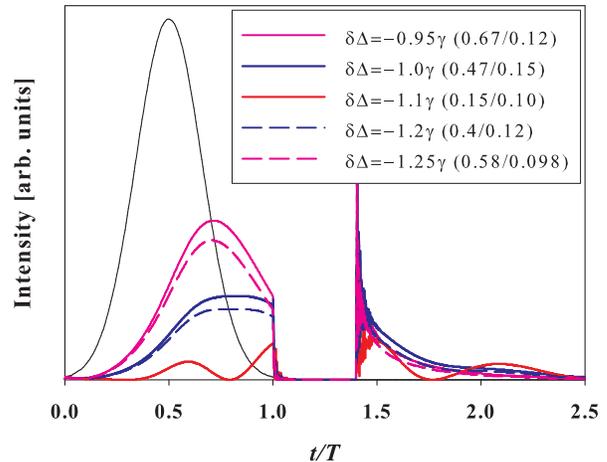}%
\caption{(Color online) Time dependence of the probe pulse propagating through
the sample when the coupling light is switched off and switched
on again after a controllable delay. The input pulse characteristics
and the sample depth are the same as in figures \ref{fig3} and
\ref{fig4}. The numbers in brackets indicate the
transmittance/retrieval efficiency with respect to the input pulse.}
\label{fig5}%
\end{figure}

In figure \ref{fig5} we demonstrate the complete memory protocol, when the
coupling light is switched off and switched on again after a controllable
delay. The input pulse characteristics and the sample depth are given by the
same values as in figures \ref{fig3} and \ref{fig4}. From the plotted graphs it
is evident that after a round of fast transient processes (poorly resolved in
time scale of these graphs) the retrieved parts of the pulses perfectly
reproduce the delayed parts of the original pulses. This can be seen by direct
comparison with figure \ref{fig4}. The transient oscillations have a frequency
around $|\Delta|\sim 50\gamma$ and a relaxation rate around $\gamma$, such that
they extend on very short time compared with the pulse duration. For storing
such a long pulse, in the considered configuration, the efficiency around $15
\%$ is achievable for effective overlap with the dispersion part of the sample
susceptibility, see figure \ref{fig3}.

\begin{figure}[t]
\includegraphics{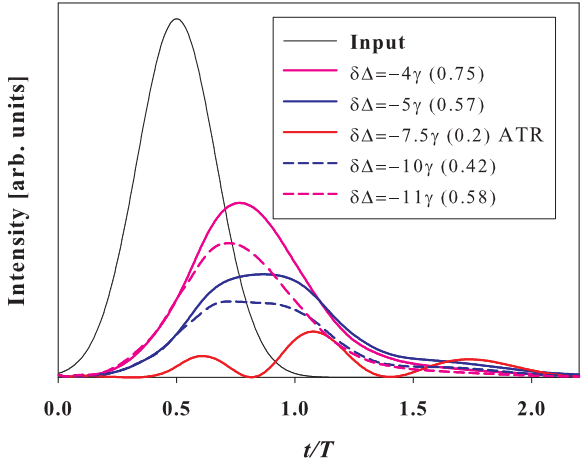}%
\caption{(Color online) Time dependence of the output pulses, when the coupling
field is on resonance with the atomic transition,
$\Delta=0$. The arrival time of the probe pulse is given by
$T/2=\gamma^{-1}$ and other parameters are the same as
in figures \ref{fig3} and \ref{fig4}: $\Delta\Omega=2\pi/T$
$\Omega_c=15\gamma$, $n_0\lambdabar^2L=25$. The spectral overlap
of the probe pulse with the lower frequency component of the
Autler-Townes doublet is controlled by mismatching the carrier
frequencies of the probe and coupling modes
$\delta\Delta=\bar{\Delta}$. The numbers in brackets indicate
the transmittance efficiency. The red curve corresponds to
coincidence of the pulse carrier frequency with the Autler-Townes
resonance and its oscillations indicate the spectral
splitting of the pulse by this resonance.}
\label{fig6}%
\end{figure}

Let us consider now how the Raman protocol can be designed for storing shorter
pulses with duration comparable to the atomic natural lifetime. In figure
\ref{fig6} we show the time dependence for a set of output probe pulses, when
the coupling field is applied exactly at resonance to the atomic transition,
such that $\Delta=0$. In this case the spectral widths of both components of
the Autler-Townes doublet are around the natural decay rate and the Raman
scheme can be applicable for delay of short probe pulses. For the graphs
plotted in figure \ref{fig6} the arrival time is given by $T/2=\gamma^{-1}$ and
for $\Delta\Omega=2\pi/T$ the duration of the pulse, described by the time
profile (\ref{4.1}), is approximately the same as the atomic natural lifetime.
Other relations and parameters are specified the same as for the graphs of
figures \ref{fig3} and \ref{fig4}. As a consequence the time dependence for the
transmitted pulses, shown in figure \ref{fig6}, for varied mismatching of their
carrier frequencies from the reference frequency of the coupling mode, look
qualitatively similar to the relevant time dependence shown in figure
\ref{fig4}. However the quantitative difference is great since the pulse
duration and delay time are much shorter in the case of figure \ref{fig6}. The
qualitative similarity in propagation of long and short pulses under stimulated
Raman protocol is explained by the similarity of the absorption/dispersion
spectral profile of the Autler-Townes resonance for $|\Delta|\gg
\Omega_c\gg\gamma$ (fig.\ref{fig4}) and for $\Delta=0$, $\Omega_c\gg\gamma$
(fig.\ref{fig6}).

\begin{figure}[t]
\includegraphics{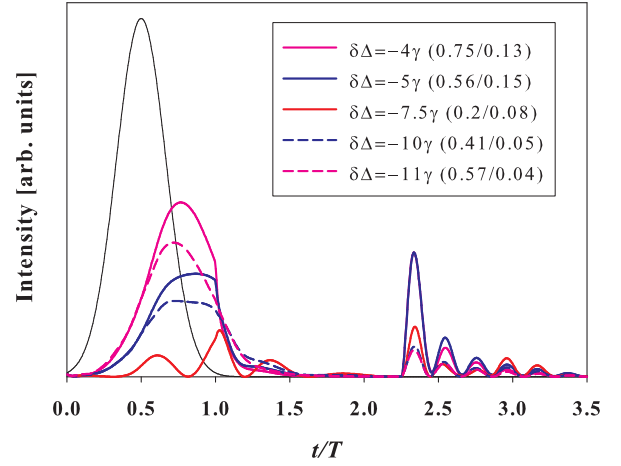}%
\caption{(Color online) Time dependence of the probe pulse propagating through
the sample when the coupling light was switched off and switched
on again after controllable delay. The input pulse characteristics
and the sample depth are the same as in figure \ref{fig6}. The
numbers in the brackets indicate the transmittance/retrieval
efficiency in respect to the input pulse. The retrieved parts of
the the probe pulses have evident manifestation of the Rabi-type
oscillations associated with the coupling field.}
\label{fig7}%
\end{figure}

The situation changes dramatically if we look at the complete memory protocol
applied to short pulses, see figure \ref{fig7}. As seen from this graph the
role of transient processes becomes much more important in this case. There are
evident manifestations of strong Rabi oscillations after switching off the
control pulse as well as after switching it on again. These Rabi oscillations
are damped on a time scale associated with the atomic natural lifetime. This
time scale coincides with the pulse duration itself, such that the Rabi
oscillations always accompany the retrieval step of the memory protocol under
the described conditions. Let us point out that the calculations presented in
figure \ref{fig7} were done for a realistic value of the atomic optical depth
and are feasible for experimental verification.

\section{Conclusion}
The presented results and our discussion did not pursue the precise
optimization of the pulse characteristics for efficient retrieval. But we can
make some general observations, concerning the applicability of the quantum
memory protocol. Depending on the type of the memory protocol it may be or may
not be sensitive to the frequency detuning of the coupling mode from the
unperturbed atomic resonance transition. For the EIT regime and in pure
$\Lambda$ configuration the probe light group velocity is not sensitive to the
location of the transparency point. If the on-resonance optical depth of the
sample is large enough the same integral efficiency for the probe pulse storage
and retrieval can be probably achieved at any detuning by proper adjustment of
the original pulse parameters. In free space configuration the probe pulse
stored under EIT protocol should preferably have duration much longer than the
atomic natural lifetime.

In contrast for the Raman protocol the pulse dynamics is differently described
under resonance and non-resonance conditions. An important consequence of our
analysis is that the Raman protocol can be applicable for storing the pulses
with duration comparable to the atomic natural lifetime. The main experimental
difficulty in achieving the natural lifetime limit is the large magnitude of
the Rabi frequency of the coupling field, which in such a case would be hard to
stabilize. As we have pointed out and supported by our numerical simulations
presented in section \ref{IV}, the EIT and Raman protocols do not compete but
actually complement one another. Each protocol has certain advantages in the
area of its applicability. From the practical point of view for both the
protocols the retrieval efficiency is expected to be high enough for sample
with on-resonance optical depth around hundreds and at such depths would be
quite tolerant to the variation of the probe pulse shape.

We have also considered the manifestation of transient processes associated
with switching off/on the control pulse. As shown by our numerical simulations
the storage of short coherent pulses via the Raman protocol is accompanied by
Rabi-type modulations of the retrieved pulse. The decay time of the Rabi-type
oscillations and the pulse duration have approximately the same order of
magnitude in this case. In the context of the memory protocol this is not a
noteworthy effect when the coherent probe pulse approaches the single photon
state and the encoded quantum information is associated with the polarization
degrees of freedom of the probe light, see Ref.\cite{CDLK}. But the effect
becomes very important for continuous variable schemes because it essentially
modifies the spectral properties of the probe light as a carrier of quantum
information. In this sense we point out that undesirable manifestation of the
transient processes is usually less important for pulses, with duration much
longer than the atomic natural lifetime. If storage and retrieval of such long
pulses is controlled by the coupling field  applied in the wing of the atomic
resonance line, then the modulation frequency of the Rabi-type oscillations is
increased and the degradation of the transient processes has much shorter time
extension than the pulse duration itself.

The above recommendations are quite general and are not constrained by unique
properties of the $\Lambda$-scheme. Let us briefly comment how the developed
approach can be further generalized to the situation of real alkali atom with
attention payed to the hyperfine structure in the upper state. It is important
to take into consideration the hyperfine effects when probing the system in the
wings of either the $D_1$ or $D_2$ lines. For the coupling field tuned exactly
in atomic resonance or detuned from it by a distance of few $\gamma$ the
$\Lambda$-type approximation seems quite realistic and our results are
applicable. For the practically important situation when the Rabi frequency is
expected to be much less than the hyperfine splitting ($\Omega_c\ll\Delta
E_{\mathrm{hpf}}$) and in the case of far detuning it is natural to expect a
generalization of the Heisenberg approach introduced in Ref.\cite{MKMP} by
adding the scattering terms and the respective Langevin forces in. The
important point for including the losses is that the incoherent scattering
should be considered not only from the atom-associated components of the
Autler-Townes structure (which in the limit $\Omega_c,\gamma\ll|\Delta|$
reproduce the atomic resonances) but also from the field-associated component
(which in the same limit is located near the field mode), see figure
\ref{fig2}. It is expected that the field-associated component will be
significantly modified due to effects of hyperfine interaction.

\section*{Acknowledgments}
We thank Prof. E.S. Polzik for fruitful discussions. The work was supported by
INTAS (project ID: 7904) and by RFBR (project \# 08-02-91355). O.S.M. and
A.S.S. would like to acknowledge the financial support from the charity
Foundation "Dynasty". O.S.M. would also like to acknowledge financial support
from the President Foundation of Russian Federation.

\appendix
\section{The atomic Green's functions and sample susceptibility}\label{A}

\subsection{Monochromatic coupling field}
The retarded-type Green's function of the excited atomic state dressed by
interaction with the vacuum modes is expressed by the standard Dyson equation
\begin{equation}
\scalebox{1.0}{\includegraphics*{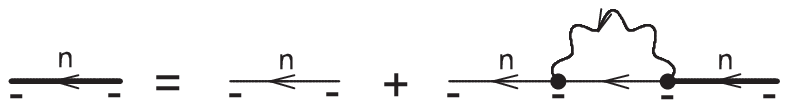}}%
\label{a.1}%
\end{equation}
and then interaction with the coupling field modifies it to the following form
\begin{equation}
\scalebox{1.0}{\includegraphics*{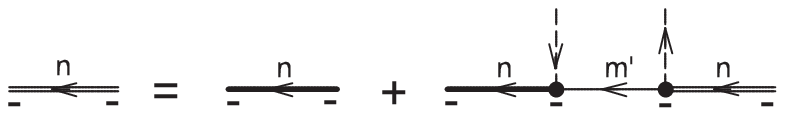}}%
\label{a.2}%
\end{equation}
The thin solid and wavy lines are respectively non-perturbed vacuum Green's
functions of the atom and field. Other notation is explained in the main text
of the paper.

For excitation of the atom by a monochromatic coupling mode these equations can
be solved by Fourier transformation. In the reciprocal space one has
\begin{eqnarray}%
G_{nn}^{(--)}(\mathbf{p},E)&=&\hbar%
\frac{E-E_{m'}(\mathbf{p},\omega)}%
{\left[E-E_{n+}(\mathbf{p},\omega)\right]%
\left[E-E_{n-}(\mathbf{p},\omega)\right]},%
\nonumber\\\nonumber\\%
E_{m'}(\mathbf{p},\omega)&=&E_{m'}+\frac{\mathbf{p}^2}{2\mathrm{m}}+%
\hbar\left(\omega-\frac{\mathbf{k p}}{\mathrm{m}}\right)%
\label{a.3}%
\end{eqnarray}%
Here $\mathbf{k}$ and $\omega=\omega_{\mathbf{k}}$ are respectively the wave
vector and frequency of the coupling mode and denominator is expressed by the
two resonance components of the Autler-Townes doublet
\begin{eqnarray}
\lefteqn{E_{n\pm}(\mathbf{p},\omega)=E_{m'}+\frac{\mathbf{p}^2}{2\mathrm{m}}%
+\frac{\hbar}{2}\left[\omega-\frac{\mathbf{k
p}}{\mathrm{m}}+\omega_{nm'}-i\frac{\gamma}{2}\right]}%
\nonumber\\%
&&\pm\left[|V_{nm'}|^2+\frac{\hbar^2}{4}%
\left(\omega_{nm'}-\omega+\frac{\mathbf{k
p}}{\mathrm{m}}-i\frac{\gamma}{2}\right)^2\right]^{1/2}%
\label{a.4}%
\end{eqnarray}
where $\omega_{nm'}=(E_n-E_{m'})/\hbar$ is the transition frequency and
$\gamma$ is the rate of natural decay of the upper state. Applying the reverse
Fourier transform over the "energy" argument $E$ one arrives to the following
Green's function, defined in the mixed time and momentum representation
\begin{eqnarray}
\lefteqn{G_{nn}^{(--)}(\mathbf{p};t,t')=-i\theta(t-t')}%
\nonumber\\%
&&\times\left[\frac{E_{n+}(\mathbf{p},\omega)-E_{m'}(\mathbf{p},\omega)}%
{E_{n+}(\mathbf{p},\omega)-E_{n-}(\mathbf{p},\omega)}\,%
\mathrm{e}^{-\frac{i}{\hbar}E_{n+}(\mathbf{p},\omega)(t-t')}\right.%
\nonumber\\%
&&\left.+\frac{E_{n-}(\mathbf{p},\omega)-E_{m'}(\mathbf{p},\omega)}%
{E_{n-}(\mathbf{p},\omega)-E_{n+}(\mathbf{p},\omega)}\,%
\mathrm{e}^{-\frac{i}{\hbar}E_{n-}(\mathbf{p},\omega)(t-t')}\right]%
\label{a.5}
\end{eqnarray}
The reverse Fourier transform from momentum to spatial coordinates
\begin{equation}
G_{nn}^{(--)}(\mathbf{r},t;\mathbf{r}',t')=\int\!\frac{d^3p}{(2\pi\hbar)^3}\,%
\mathrm{e}^{\frac{i}{\hbar}\mathbf{p}(\mathbf{r}-\mathbf{r}')}%
G_{nn}^{(--)}(\mathbf{p};t,t')%
\label{a.6}%
\end{equation}
cannot be evaluated analytically in the general case. However when considering
ultracold atoms with temperature significantly lower than the Doppler limit the
effect of atomic motion is rather weak and can be taken into account as a small
correction to the approximation of stationary atoms. For the Raman process,
when the frequency detuning between the coupling mode and atomic frequency is
large enough, the effect of atomic motion becomes even less important in the
calculation of the sample susceptibility. But it gives the principle limitation
for the storage time of the memory protocol as discussed in subsection
\ref{IIIB} and commented below in sec. \ref{A4}.

\subsection{Coupling with a rectangular time envelope profile}

If the action of the coupling mode is limited in time and described by a
rectangular envelope profile (\ref{3.9}) inside the interval $0<t<T$, then the
Green's function $G_{nn}^{(--)}(\mathbf{p};t,t')$ defined in the mixed
representation for different relations between times $t$ and $t'$ is given by

\noindent 1.1) $t'<0$ and $t<0$
\begin{eqnarray}
G_{nn}^{(--)}(\mathbf{p};t,t')&=&-i\theta(t-t')%
\mathrm{e}^{-\frac{i}{\hbar}E_n(\mathbf{p})(t-t')-\frac{\gamma}{2}(t-t')}
\nonumber\\%
E_n(\mathbf{p})&=&E_n+\frac{\mathbf{p}^2}{2\mathrm{m}}%
\label{a.7}%
\end{eqnarray}
1.2) $t'<0$ and $0<t<T$
\begin{eqnarray}
\lefteqn{G_{nn}^{(--)}(\mathbf{p};t,t')=}%
\nonumber\\%
&&-i\left[\frac{E_{n+}(\mathbf{p},\omega)-E_{m'}(\mathbf{p},\omega)}%
{E_{n+}(\mathbf{p},\omega)-E_{n-}(\mathbf{p},\omega)}\,%
\mathrm{e}^{-\frac{i}{\hbar}E_{n+}(\mathbf{p},\omega)t}\right.%
\nonumber\\%
&&\left.+\frac{E_{n-}(\mathbf{p},\omega)-E_{m'}(\mathbf{p},\omega)}%
{E_{n-}(\mathbf{p},\omega)-E_{n+}(\mathbf{p},\omega)}\,%
\mathrm{e}^{-\frac{i}{\hbar}E_{n-}(\mathbf{p},\omega)t}\right]%
\nonumber\\ \nonumber\\%
&&\times\mathrm{e}^{+\frac{i}{\hbar}E_n(\mathbf{p})t'+\frac{\gamma}{2}t'}%
\label{a.8}%
\end{eqnarray}%
1.3) $t'<0$ and $T<t$
\begin{eqnarray}
\lefteqn{G_{nn}^{(--)}(\mathbf{p};t,t')=-i%
\mathrm{e}^{-\frac{i}{\hbar}E_n(\mathbf{p})(t-T)-\frac{\gamma}{2}(t-T)}}%
\nonumber\\%
&&\left[\frac{E_{n+}(\mathbf{p},\omega)-E_{m'}(\mathbf{p},\omega)}%
{E_{n+}(\mathbf{p},\omega)-E_{n-}(\mathbf{p},\omega)}\,%
\mathrm{e}^{-\frac{i}{\hbar}E_{n+}(\mathbf{p},\omega)T}\right.%
\nonumber\\%
&&\left.+\frac{E_{n-}(\mathbf{p},\omega)-E_{m'}(\mathbf{p},\omega)}%
{E_{n-}(\mathbf{p},\omega)-E_{n+}(\mathbf{p},\omega)}\,%
\mathrm{e}^{-\frac{i}{\hbar}E_{n-}(\mathbf{p},\omega)T}\right]%
\nonumber\\ \nonumber\\%
&&\times\mathrm{e}^{+\frac{i}{\hbar}E_n(\mathbf{p})t'+\frac{\gamma}{2}t'}%
\label{a.9}%
\end{eqnarray}%
2.1) $0<t'<T$ and $0<t<T$
\begin{eqnarray}
\lefteqn{G_{nn}^{(--)}(\mathbf{p};t,t')=-i\theta(t-t')}%
\nonumber\\%
&&\times\left[\frac{E_{n+}(\mathbf{p},\omega)-E_{m'}(\mathbf{p},\omega)}%
{E_{n+}(\mathbf{p},\omega)-E_{n-}(\mathbf{p},\omega)}\,%
\mathrm{e}^{-\frac{i}{\hbar}E_{n+}(\mathbf{p},\omega)(t-t')}\right.%
\nonumber\\%
&&\left.+\frac{E_{n-}(\mathbf{p},\omega)-E_{m'}(\mathbf{p},\omega)}%
{E_{n-}(\mathbf{p},\omega)-E_{n+}(\mathbf{p},\omega)}\,%
\mathrm{e}^{-\frac{i}{\hbar}E_{n-}(\mathbf{p},\omega)(t-t')}\right]\phantom{a10}%
\label{a.10}%
\end{eqnarray}%
2.2) $0<t'<T$ and $T<t$
\begin{eqnarray}
\lefteqn{G_{nn}^{(--)}(\mathbf{p};t,t')=-i%
\mathrm{e}^{-\frac{i}{\hbar}E_n(\mathbf{p})(t-T)-\frac{\gamma}{2}(t-T)}}%
\nonumber\\%
&&\left[\frac{E_{n+}(\mathbf{p},\omega)-E_{m'}(\mathbf{p},\omega)}%
{E_{n+}(\mathbf{p},\omega)-E_{n-}(\mathbf{p},\omega)}\,%
\mathrm{e}^{-\frac{i}{\hbar}E_{n+}(\mathbf{p},\omega)(T-t')}\right.%
\nonumber\\%
&&\left.+\frac{E_{n-}(\mathbf{p},\omega)-E_{m'}(\mathbf{p},\omega)}%
{E_{n-}(\mathbf{p},\omega)-E_{n+}(\mathbf{p},\omega)}\,%
\mathrm{e}^{-\frac{i}{\hbar}E_{n-}(\mathbf{p},\omega)(T-t')}\right]\phantom{a11}%
\label{a.11}%
\end{eqnarray}%
3) $T<t'$ and $T<t$
\begin{equation}
G_{nn}^{(--)}(\mathbf{p};t,t')=-i\theta(t-t')%
\mathrm{e}^{-\frac{i}{\hbar}E_n(\mathbf{p})(t-t')-\frac{\gamma}{2}(t-t')}
\label{a.12}%
\end{equation}

\noindent These expressions for the Green's function in different time domains
can be built up by clipping the partial solutions for different excitation
regimes. These include the atomic dynamics driven by either interaction only
with the vacuum modes (\ref{a.7}), (\ref{a.12}) or jointly with the vacuum and
coupling modes (\ref{a.5}). The required boundary conditions are dictated by
continuity at the points $t=0,T$.

\subsection{Susceptibility of the sample in response to the probe field}

The susceptibility of the atomic sample in response to the probe field is given
by the polarization operator contributing to the Dyson equation (\ref{3.1})
\begin{eqnarray}
\lefteqn{\chi(\mathbf{r}_{\bot},z;t,t')=%
-\frac{1}{\hbar}\left|\left(\mathbf{d e}_{-}\right)_{nm}\right|^2}%
\nonumber\\%
&&\times\int\frac{d^{3}p}{(2\pi\hbar)^3}n_0(\mathbf{r}_{\bot},z)%
f_0(\mathbf{p}_{\bot},p_z-\hbar\bar{k})%
\nonumber\\%
&&\times\mathrm{e}^{-\frac{i}{\hbar}E_m(\mathbf{p}_{\bot},p_z-\hbar\bar{k})(t'-t)}%
G_{nn}^{(--)}(\mathbf{p};t,t')%
\label{a.13}%
\end{eqnarray}
where $n_0(\mathbf{r}_{\bot},z)$ is the spatial distribution of atomic density.
The total energy of the atom in the ground state
\begin{equation}
E_m(\mathbf{p}_{\bot},p_z-\hbar\bar{k})=E_m+\frac{\mathbf{p}_{\bot}^2}{2\mathrm{m}}+%
\frac{(p_{z}-\hbar\bar{k})^2}{2\mathrm{m}}%
\label{a.14}%
\end{equation}
is described by the kinetic part, written as function of atomic momentum
$\mathbf{p}=(\mathbf{p}_{\bot},p_z)$ of the excited state and shifted by the
momentum of the photon from the probe mode $\hbar\bar{k}$. Such a momentum
conservation in the coherent process (\ref{3.1}) is also taken into account in
the argument of the Maxwell distribution
$f_0(\mathbf{p}_{\bot},p_z-\hbar\bar{k})$.

It is convenient to define the slow-varying component of the
susceptibility
\begin{equation}
\tilde{\chi}(\mathbf{r}_{\bot},z;t,t')=\mathrm{e}^{i\bar{\omega}(t-t')}%
\chi(\mathbf{r}_{\bot},z;t,t')%
\label{a.15}%
\end{equation}
which describes the dynamics of the slow-varying amplitude of the probe field,
see Eqs.(\ref{3.3}), (\ref{3.6}), (\ref{3.12}), (\ref{3.14}).

\subsection{The Green's function and sample
susceptibility modified by interruption of the control
pulse}\label{A4}

In the general case the retarded-type Green's function responsible for atomic
dynamics, dressed by two pulses of the coupling field (see diagram
(\ref{3.11})), has a rather cumbersome analytical structure. The result can be
simplified for the most important situation when time $t'$ belongs to the first
control pulse and time $t$ belongs to the second pulse. Let us denote the delay
between the two pulses by $T_D$ and superscribe all parameters associated with
the second pulse, such as duration, Rabi frequency, transition matrix elements
etc., by a prime. Then for $0<t'<T$ and $T+T_D<t<T+T_D+T'$ the Green's
function, expressed by diagram (\ref{3.11}), is given by
\begin{eqnarray}
\lefteqn{G_{n/\!/n}^{(--)}(\mathbf{p};t,t')=-i%
\frac{V'_{nm'}V_{nm'}^{*}}{(E'_{n+}-E'_{n-})(E_{n+}-E_{n-})}}%
\nonumber\\ \nonumber\\%
&&\times\left[\mathrm{e}^{-\frac{i}{\hbar}E'_{n+}(t-T-T_D)}-%
\mathrm{e}^{-\frac{i}{\hbar}E'_{n-}(t-T-T_D)}\right]%
\mathrm{e}^{-\frac{i}{\hbar}E_{m'}(\mathbf{p},\omega)T_D}%
\nonumber\\ \nonumber\\%
&&\times\left[\mathrm{e}^{-\frac{i}{\hbar}E_{n+}(T-t')}-%
\mathrm{e}^{-\frac{i}{\hbar}E_{n-}(T-t')}\right]%
\label{a.16}%
\end{eqnarray}
were the matrtix element $V'_{nm'}$ is related to the retrieval coupling field.
The susceptibility component responsible for the recovering process, described
by the first diagram on the right hand side of Eq. (\ref{3.10}), is given by
\begin{eqnarray}
\lefteqn{\chi_{/\!/}(\mathbf{r}_{\bot},z;t,t')=%
-\frac{1}{\hbar}\left|\left(\mathbf{d e}_{-}\right)_{nm}\right|^2}%
\nonumber\\%
&&\times\int\frac{d^{3}p}{(2\pi\hbar)^3}n_0(\mathbf{r}_{\bot},z)%
f_0(\mathbf{p}_{\bot},p_z-\hbar\bar{k})%
\nonumber\\%
&&\times\mathrm{e}^{-\frac{i}{\hbar}E_m(\mathbf{p}_{\bot},p_z-\hbar\bar{k})(t'-t)}%
G_{n/\!/n}^{(--)}(\mathbf{p};t,t')%
\label{a.17}%
\end{eqnarray}
and its slow-varying component $\tilde{\chi}_{/\!/}(\mathbf{r}_{\bot},z;t,t')$,
contributing to Eq.(3.12), is defined by the transformation (\ref{a.15}).

It is important to recognize that validity of expression (\ref{a.17}) is
restricted by the global effect of atomic motion for the entire memory
protocol, which includes both the storage and retrieval stages. In our
derivation of the susceptibility (\ref{a.17}), driven by two control pulses, we
have assumed that during the delay time $T_D$ atoms do not noticeably change
their location on the scale of the spatial inhomogeneity associated with their
original density distribution $n_0(\mathbf{r}_{\bot},z)$.


\begin{thebibliography}{99}
\bibitem{Coherence} More precisely, in the case of atomic subsytem we require
    that its behavior is completely described by the single particle density matrix, see below.
\end{thebibliography}
\end{document}